\documentclass[final,5p,times,twocolumn]{elsarticle}

\usepackage{amsmath, amssymb}
\usepackage{lipsum}
\journal{Results in Physics}
\usepackage{graphicx}
\usepackage[caption=false]{subfig}
\usepackage{dcolumn}
\usepackage{bm}
\usepackage{natbib}
\usepackage{url}

\usepackage{tikz}
\usetikzlibrary{shapes}
\usetikzlibrary{positioning}
\usetikzlibrary{calc} 
\usepackage{braket}
\usepackage{bbm}
\usepackage{esvect}
\usepackage{ulem}
\usepackage[colorlinks=true,backref=false, linktocpage=true,
citecolor=blue,urlcolor=blue,linkcolor=blue,pdfpagemode=UseOutlines]{hyperref}
\usepackage{multirow}
\usepackage[section]{placeins}
\usepackage{booktabs}
\usepackage{appendix}
\usepackage{empheq}

\begin{document}

\begin{frontmatter}

\title{Probing Entanglement Dynamics in the SYK Model using Quantum Computers}

\author[first,second]{Talal Ahmed Chowdhury}
\address[first]{Department of Physics, University of Dhaka, P.O. Box 1000, Dhaka, Bangladesh.}
\address[second]{Department of Physics and Astronomy, University of Kansas, Lawrence, Kansas 66045, USA.}
\ead{talal@ku.edu}
\author[third]{Kwangmin Yu}
\address[third]{Computational Science Initiative, Brookhaven National Laboratory, Upton, New York 11973, USA.}
\ead{kyu@bnl.gov}
\author[fifth,sixth,seventh]{Raza Sabbir Sufian}
\address[fifth]{Department of Physics, New Mexico State University, Las Cruces, NM 88003, USA}
\address[sixth]{RIKEN-BNL Research Center, Brookhaven National Laboratory, Upton, New York 11973, USA.}
\address[seventh]{Physics Department, Brookhaven National Laboratory, Upton, New York 11973, USA.}
\ead{gluon2025@gmail.com}

\begin{abstract}
Quantum computers are expected to be vital for exploring complex dynamics in many-body quantum systems. Thus, validating established results on current quantum computers is essential for evaluating their future utility. Hence, we investigate the entanglement entropy of the Sachdev-Ye-Kitaev (SYK) model, a paradigmatic model of quantum chaos, many-body physics, and holographic duality, in current IBM's superconducting quantum computers. We implement optimized swap-based many-body interference protocol and randomized measurement protocol tailored for IBM quantum computers' limited qubit connectivity. Additionally, we employ quantum multi-programming that parallelizes circuit execution to improve the results obtained by the randomized measurement protocol. Finally, by incorporating the quantum error mitigation techniques into our implementation of the entropy measurement protocols on IBM quantum hardware, we show that the current noisy quantum computer can yield results aligned with theoretical expectations, therefore affirming its capability to explore chaotic quantum dynamics in complex quantum systems.
\end{abstract}

\end{frontmatter}

\section{Introduction}\label{sec:introduction}
Quantum entanglement~\cite{horodecki} plays a crucial role in various fields, including quantum computing, quantum information science, many-body physics, high-energy and nuclear physics, and black hole physics. Additionally, effectively harnessing quantum information and entanglement dynamics is essential for advancing quantum technologies and exploring complex quantum systems~\cite{Jozsa-entanglement, Preskill-entanglement}. However, the exponential growth of the Hilbert space associated with increasing system size poses significant challenges for classical computers, making it difficult to compute entanglement entropy for highly entangled states. Besides, experimentally measuring the entanglement entropy and its dynamics in complex quantum systems is challenging due to the specific requirements for state preparation, the execution of time evolution, and the implementation of entropy measurement techniques, such as those based on quantum gas microscopy~\cite{quantum-gas-microscope-1, quantum-gas-microscope-2}.

One such complex theoretical model for quantum man-body system is the  Sachdev-Ye-Kitaev (SYK) model~\cite{Sachdev-Ye, kitaev, polchinski,maldacena}, a $0+1$ dimensional quantum mechanical system consisting of $N$ Majorana fermions with random interactions involving $q$ of these fermions at a time, where $q$ is an even number. It is notable for its exact solvability at strong coupling, maximal chaos, and emergent conformal symmetry. Furthermore, the SYK model is a prime example of a strongly interacting quantum non-Fermi liquid lacking a quasiparticle description (for a detailed review of the SYK Model, refer to Ref.~\cite{chowdhury-syk-rmp}). Quantum entanglement plays a crucial role in revealing the nontrivial correlations among the constituents of this representative model~\cite{Liu-SYK,Zhang-SYK}. Thus, given the SYK model's rich physics and the rarity of such solvable, strongly interacting chaotic systems in quantum mechanics, experimentally measuring its quantum entanglement is a valuable endeavor.

In this regard, quantum computers can be utilized as experimental probes to measure the entanglement entropy of the SYK model. In fact, real quantum devices are emerging as highly effective programmable quantum platforms, capable of accurately preparing complex quantum many-body models, and offer a valuable testing ground for novel phenomena that would be difficult or impossible to realize through experimental means~\cite{Swan-dynamics-quantum-info-review, Georgescu:2013oza, Daley:2022eja}. Programmable quantum platforms are being used to study discrete time crystals~\cite{Zhang-time-crystal, Randall-MBL-time-crystal}, many-body localization~\cite{Schreiber-MBL, Choi-MBL, Bernien-MBL, Zhang-MBL}, quantum phases, and topologically ordered states~\cite{Mazurenko-Quantum-phase, Scholl-Quantum-phase, Ebadi-Quantum-phase, Satzinger-topological-ordered-state, Dumitrescu-topological-phase, Google-topological-order, Iqbal-topological-order} in quantum many-body systems. Building on our previous work that investigated the feasibility of current quantum computers as programmable platform to study the Page Curve entanglement dyanicms~\cite{page-curve-paper}, we now shift our focus to a more complex measurement of entanglement dynamics in the SYK model. Although some studies~\cite{alvarez-syk, Luo-syk, Danshita-SYK, Pikulin-SYK, chew-SYK, Chen-SYK, Babbush-SYK, Jafferis:2022crx, kobrin, Raghav-SYK} have addressed the time-evolution of the SYK model in quantum computers, we employ an optimized version of swap-based measurement protocol, delineated in~\cite{page-curve-paper}, and randomized measurement protocols~~\cite{Brydges-randomized, Enk-Beenakker, Elben-1, Vermersch-1, Elben-2, Rath, Elben-3} to probe the the entanglement entropy within this model using IBM's superconducting quantum computer. Specifically, we study the growth of entanglement entropy for subsystems of varying sizes over time by evolving a total product state using the time evolution operator associated with the SYK Hamiltonian. 

However, implementing any quantum protocols on current quantum devices, such as IBM's quantum processors, presents significant challenges due to the errors and noise inherent in these systems. To address this, we apply various quantum error mitigation (QEM) techniques that have been developed and demonstrate their effectiveness in practical applications~\cite{Kim-error-mitigation, kim2023evidence, charles2305simulating, chowdhury2024enhancing}. As a result, we accurately measure the variation of entanglement entropy over time during the evolution under the SYK Hamiltonian in quantum computers, even for long circuit depths exceeding 600.

The structure of the paper is as follows. In section~\ref{sec:SYK model}, we provide a telegraphic description of the SYK model and the entanglement entropy measure, namely the second R\'enyi entropy that we compute in IBM's superconducting quantum computers. Section~\ref{sec:entanglement} describes the two entanglement entropy measuring protocols adopted in this work. In section~\ref{sec:IBMQ}, we present our results of the variation of the entanglement entropy with time under the time evolution by the SYK Hamiltonian, which we have obtained by employing the entanglement measuring protocols in combination with quantum error mitigation on IBM's quantum devices. Finally, we conclude in section~\ref{sec:conclusion}.

\section{The SYK Model}\label{sec:SYK model}
The SYK model is given by the Hamiltonian,
\begin{equation}
    H = i^{\frac{q}{2}}\sum_{1\leq i_{1}<i_{2}<\cdots <i_{q}\leq N}J_{i_{1}i_{2}\cdots i_{q}}\,\chi_{i_{1}}\chi_{i_{2}}\cdots \chi_{i_{q}},
    \label{eq:SYK-Hamiltonian}
\end{equation}
where $\chi_{i}$ are the Majorana fermions. Each coefficient $J_{i_{1}i_{2}\cdots i_{q}}$ is taken as a real number drawn from a random Gaussian distribution such that $\langle J_{i_{1}i_{2} \cdots i_{q}}\rangle = 0$ and $\langle J^{2}_{i_{1}i_{2}\cdots i_{q}}\rangle = \frac{(q-1)!\mathcal{J}^2}{N^{q-1}}$. In this study, we restrict ourselves to $q=4$ as this case represents the dominant low-energy model with time-reversal symmetric interactions.

To implement the SYK Hamiltonian in quantum computers, we use the Jordan-Wigner transformation~\cite{jordan-wigner} to map the Majorana fermions into Pauli operators in the following way,
\begin{equation}
    \chi_{2i-1}\rightarrow Z_{1}Z_{2}\cdots Z_{i-1}X_{i},\,\,\,
    \chi_{2i}\rightarrow Z_{1}Z_{2}\cdots Z_{i-1}Y_{i}.
    \label{eq:JW-transform}
\end{equation}
The Hamiltonian now becomes,
\begin{equation}
    H = \sum_{n=1}^{Q}a_{n}\hat{P}_{n},
    \label{eq:SYK-pauli-terms}
\end{equation}
where $Q$ is the number of terms in Eq. (\ref{eq:SYK-Hamiltonian}), $\hat{P}_{n}$ is the Pauli operator of weight $N/2$, i.e. it acts on the Hilbert space of $N/2$ qubits, and $a_{n}$ is the corresponding coefficient.

Here, we introduce the entanglement entropy measure we use in this study. Consider a system of $N$ qubits in a pure state $|\psi\rangle$, with its density matrix given by $\rho = |\psi\rangle\langle\psi|$. If we partition this system into two subsystems consisting of  $L$ and $M = N - L$ qubits, the entanglement entropy, specifically the  $n$-R\'enyi entropy associated with the $L$-qubit subsystem, is defined as
\begin{equation}
S^{(n)}_{L} = \frac{1}{1-n} \log \left[ \mathrm{Tr}(\rho_{L}^{n}) \right],
\label{eq:n-Renyi entropy}
\end{equation}
where $\rho_{L}$ is the reduced density matrix for the subsystem with  $L$ qubits. This reduced density matrix is obtained by tracing out the subsystem of $M = N - L$ qubits, such that $\rho_{L} = \mathrm{Tr}_{M} \rho$. Our focus in this study is $n = 2$, corresponding to the 2-R\'enyi entropy (commonly referred to as R\'enyi entropy), $S^{(2)}$. Additionally, as $n$ approaches 1, the $n$-R\'enyi entropy converges to the von Neumann entropy:
\begin{equation}
S^{vN}_{L} = -\mathrm{Tr} \left( \rho_{L} \log \rho_{L} \right).
\label{eq:von neumann entropy}
\end{equation}

To execute the time evolution driven by the SYK Hamiltonian in quantum computers, we use first-order Trotterization~\cite{trotter, suzuki1, suzuki2},
\begin{equation}
    U(t)=e^{-i H t}\approx \left(\prod_{n=1}^{Q}e^{-i H_{n}t/r} \right)^{r},
    \label{eq:first-trotterization}
\end{equation}
where $H_{n}=a_{n}\hat{P}_{n}$ and $r$ is the number of Trotter steps. The Trotter timestep size is defined as $\delta t = t/r$. We want to emphasize that the SYK Hamiltonian includes all-to-all quartic interactions, which causes the number of terms in the Hamiltonian to grow approximately as $O(N^4)$ with respect to the number of fermions, $N$. Furthermore, the circuit depth associated with a single Trotter step in first-order Trotterization increases polynomially with $N$. For example, for $N = 6$, 8, and 10, the circuit depth $d_{\mathrm{circuit}}$ for a single Trotter step is $d_{\mathrm{circuit}} = 84$, 494, and 1726, respectively, before transpilation. Additionally, the gate complexity associated with first-order Trotterization, as shown in Refs.~\cite{alvarez-syk}, roughly scales as $O(N^{10})$, unless non-trivial optimizations are implemented (for recent works on higher-order Trotterization, see Refs~\cite{Chen-Brandao, Chen-higher-order}). Consequently, for current quantum computers, where Trotterization is the most appropriate technique for Hamiltonian simulation, the time evolution driven by the SYK Hamiltonian for larger fermion numbers $N > 6$ (or $N_{\mathrm{qubit}} > 3$) becomes highly complex.

\section{Entanglement Entropy Measuring Protocols}\label{sec:entanglement}

We consider two protocols to measure the purity and R\'enyi entropy of subsystems in a total quantum state: 1. swap-based many-body interference (SWAP-MBI) protocol and 2. randomized measurement (RM) protocol.

\begin{figure}[htbp]
\centering
\includegraphics[width=0.48\textwidth]{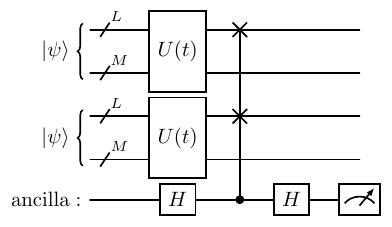}
\caption{The quantum circuit associated with the swap-based many-body interference (SWAP-MBI) protocol. Here the R\'enyi entropy of $L$-qubit subsystem is determined by executing the swap test.}
\label{fig:swap-circuit}
\end{figure}

The swap-based many-body interference protocol~\cite{Buhrman:2001rma, Horodecki-swap, Ekert-swap} is based on the swap test which is presented in the quantum circuit Fig.~\ref{fig:swap-circuit},
where the purity associated with the $L$-qubit subsystem $\mathrm{Tr}(\rho_{L}^{2})$ can be estimated by $\mathrm{Tr}(\rho_{L}^{2})=2 P_{0}-1$ where $P_{0}$ is the probability of finding the ancilla qubit in $|0\rangle$ state. Now, the R\'enyi entropy $S^{(2)}$ can be readily obtained by $S^{(2)}=-\mathrm{log}(2P_{0}-1)$.

\begin{figure}[htbp]
\centering
\includegraphics[width=0.48\textwidth]{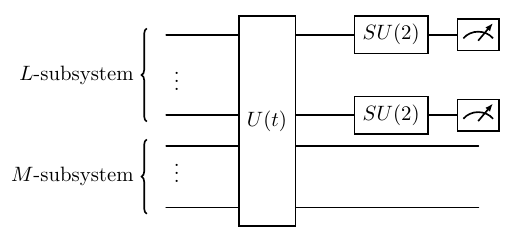}
\caption{The quantum circuit associated with the randomized measurement (RM) protocol. Here the local Haar-random unitaries $\hat{U}_{a}$ are applied at $L$-qubit subsystem whose R\'enyi entropy is being measured.}
\label{fig:randomized-circuit}
\end{figure}

On the other hand, the randomized measurement protocol (RM protocol)~\cite{Brydges-randomized, Enk-Beenakker, Elben-1, Vermersch-1, Elben-2, Rath, Elben-3} relies on the principle that the R\'enyi entropy of a quantum system is reflected in the statistical correlations between the outcomes of measurements conducted on random bases. After evolving the initial state of $N_{\mathrm{qubit}}$ system with the time evolution operator $U(t)$ to obtain the quantum state $|\psi\rangle$ or the associated density matrix $\rho=|\psi\rangle\langle\psi|$, we focus on measuring the purity and R\'enyi entropy associated with the subsystem $A$ of $L$ qubits. Now, one applies a product of single-qubit unitary operators 
\begin{equation}
 \hat{U}_{a}=U^{(2)}_{1}\otimes U^{(2)}_{2}\otimes...\otimes U^{(2)}_{L},
 \label{eq:single-qubit-local-unitary}
\end{equation}
where each of $U^{(2)}_{i}$ is drawn independently from the circular unitary ensemble (CUE) of the $SU(2)$. Afterward, the measurements of the qubits are done on the computational basis ($Z$-basis). The corresponding quantum circuit is shown in Fig.~\ref{fig:randomized-circuit}.
For each $\hat{U}_{a}$, one conducts repeated measurements to obtain the statistics for estimating the occupation probabilities, $P_{\hat{U}_{a}}(\mathbf{s}_{L})=\mathrm{Tr}[\hat{U}_{a}\rho\hat{U}^{\dagger}_{a}|\mathbf{s}_{L}\rangle\langle \mathbf{s}_{L}|]$ of computational basis states $|\mathbf{s}_{L}\rangle = |s_{1},s_{2},...,s_{L}\rangle$ with $s_{i}=0,1$. Here, $\hat{U}_{a}$ acts only on the subspace of $L$ qubits. Afterward, the entire process is repeated for $N_{U}$ different randomly drawn instances of $\hat{U}_{a}$.

After determining the set of outcome probabilities $P_{\hat{U}_{a}}(\mathbf{s}_{L})$ of the computational basis states $|\mathbf{s}_{L}\rangle$ for one instance of random unitaries, $\hat{U}_{a}$, one computes the following quantity,
\begin{equation}
    X_{a} = 2^{L}\sum_{\mathbf{s}_{L},\mathbf{s}'_{L}}(-2)^{-D[\mathbf{s}_{L},\mathbf{s}'_{L}]}P_{\hat{U}_{a}}(\mathbf{s}_{L})P_{\hat{U}_{a}}(\mathbf{s}'_{L})
\end{equation}
where $D[\mathbf{s}_{L},\mathbf{s}'_{L}]$ is the Hamming distance between bitstrings $\mathbf{s}_{L}=s_{1}s_{2}...s_{L}$ and $\mathbf{s}'_{L}=s'_{1}s'_{2}...s'_{L}$, measuring how different two bitstrings are, i. e., $D[\mathbf{s}_{L},\mathbf{s}'_{L}]\equiv \#\{i\in A|s_i\neq s'_i\}$. Consequently the ensemble average of $X_{a}$, denoted by $\overline{X}$,
\begin{equation}
    \overline{X} = \frac{1}{N_{U}}\sum_{a=1}^{N_{U}}X_{a}
    \label{eq:RM-estimated-purity}
\end{equation}
is the second-order cross-correlations across the ensemble of discrete $N_{U}$ random unitaries $\hat{U}_{a}$, and provides the estimation of the purity $\mathrm{Tr}(\rho_{L}^{2})$ associated with a subsystem of $L$ qubits. Finally, the R\'enyi entropy is 
\begin{equation}
    S^{(2)} = -\mathrm{log}\overline{X}
    \label{eq:renyi-RMP}
\end{equation}

\section{IBMQ Results}\label{sec:IBMQ}

Despite the importance of the SYK model and the entanglement entropy measurement, implementing them on near-term quantum computers is challenging due to the complexity of the Pauli strings and the computation overhead of the two entanglement entropy measurement protocols, the SWAP-MBI and the RM protocols. In particular, one Trotter step has 126 circuit depths and 39 two-qubit gates (\texttt{CX} or \texttt{CZ}). Since we compute up to five Trotter steps, the largest circuit depth and the number of two-qubit gates are 630 and 195, respectively. Based on the Trotter step implementation, the SWAP-MBI protocol needs 196 additional circuit depth to implement the SWAP test in the protocol.

On the other hand, the RM protocol needs only 5 additional circuit depths. However, the protocol needs a large number of repetitions for the random sampling of the $SU(2)$ circuits. To overcome those challenges, we apply several methods detailed below.

\begin{figure*}[h!]
	\centerline{\includegraphics[width = 1.0\textwidth]{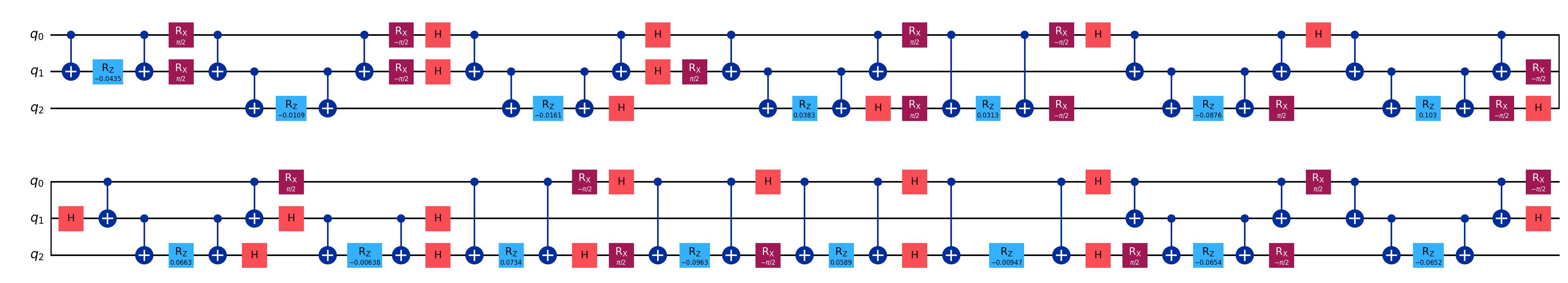}}
	\caption{The quantum circuit associated with the single Trotter step for $N=6$ Majorana fermions ($N_{\mathrm{qubit}}=3$) for a single instance of SYK Hamiltonian. The circuit diagram is generated by \texttt{qiskit.visualization.circuit$\_$drawer}. The circuit is continued from the upper panel to the lower panel.}
	\label{fig:single_trotter}
\end{figure*}

\begin{figure}[htbp]
\centering
\includegraphics[width=0.48\textwidth]{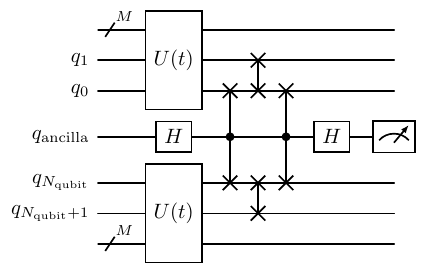}
\caption{The optimized SWAP-MBI protocol devised for IBM quantum devices with limited qubit connectivity. Here, the quantum circuit for determining the R\'enyi entropy for the $L=2$ left-subsystem, with $M=N_{\mathrm{qubit}}-2$, is shown as an example.}
\label{fig:optimized-swap-circuit}
\end{figure}

We conduct our experiment on the IBM Quantum computer, \texttt{ibm\_marrakesh} which has 156 qubits with basis gates \texttt{CZ, ID, RX, RZ, RZZ, SX, X}. We evolve a product state of $N_{\mathrm{qubit}}=3$ that corresponds to $N=6$ Majorana fermions, with time-evolution operator driven by the SYK Hamiltonian via the first-order Trotterization up to $t = 10$ (in unit of $\hbar/E$ where $E$ is the overall energy scale of the SYK Hamiltonian) in five Trotter steps (Trotter step size 2.0). We also present the quantum circuit associated with the single Trotter step in Fig.~\ref{fig:single_trotter}. In each Trotter step, we employ the entanglement entropy measuring protocols: SWAP-MBI and RM protocols to measure the entropy of left single-qubit and two-qubit subsystems, respectively. As going beyond the case $N>6$ for the SYK model is highly challenging on near-term quantum computers due to the polynomial growth of gate complexity with $N$, we focus on the feasible case of $N=6$ Majorana fermions or $N_{\mathrm{qubit}}=3$ qubits. Additionally, we employed an optimized version of the SWAP-MBI protocol, designed specifically for quantum devices with limited qubit connectivity~\cite{page-curve-paper} as shown in Fig.~\ref{fig:optimized-swap-circuit}.

Still, running quantum algorithms on near-term quantum devices, such as IBM Quantum processors, poses a significant challenge due to the errors and noise present in these systems. However, we precisely measure the variation of entanglement entropy with time under the SYK Hamiltonian evolution in quantum computers up to long circuit depths (more than 600 depths, five timesteps) by applying four quantum error mitigation methods to mitigate the errors and noise inherent to quantum devices: Zero-Noise Extrapolation (ZNE), Pauli Twirling (PT), Dynamic Decoupling (DD), and Matrix-free Measurement Mitigation (M3). Additionally, we conduct a comprehensive performance comparison of ZNE, PT, and DD for the Hamiltonian simulation on IBM quantum computers, as discussed in~\cite{Choi:2025bdw}, where we show that combining these Quantum Error Mitigation (QEM) methods, i.e., ZNE, PT, and DD, yields the highest accuracy. For more details on the QEM implementations, please refer to Refs. \cite{chowdhury2024enhancing, Choi:2025bdw}.

\begin{figure}
    \centering
    \includegraphics[width=1.0\linewidth]{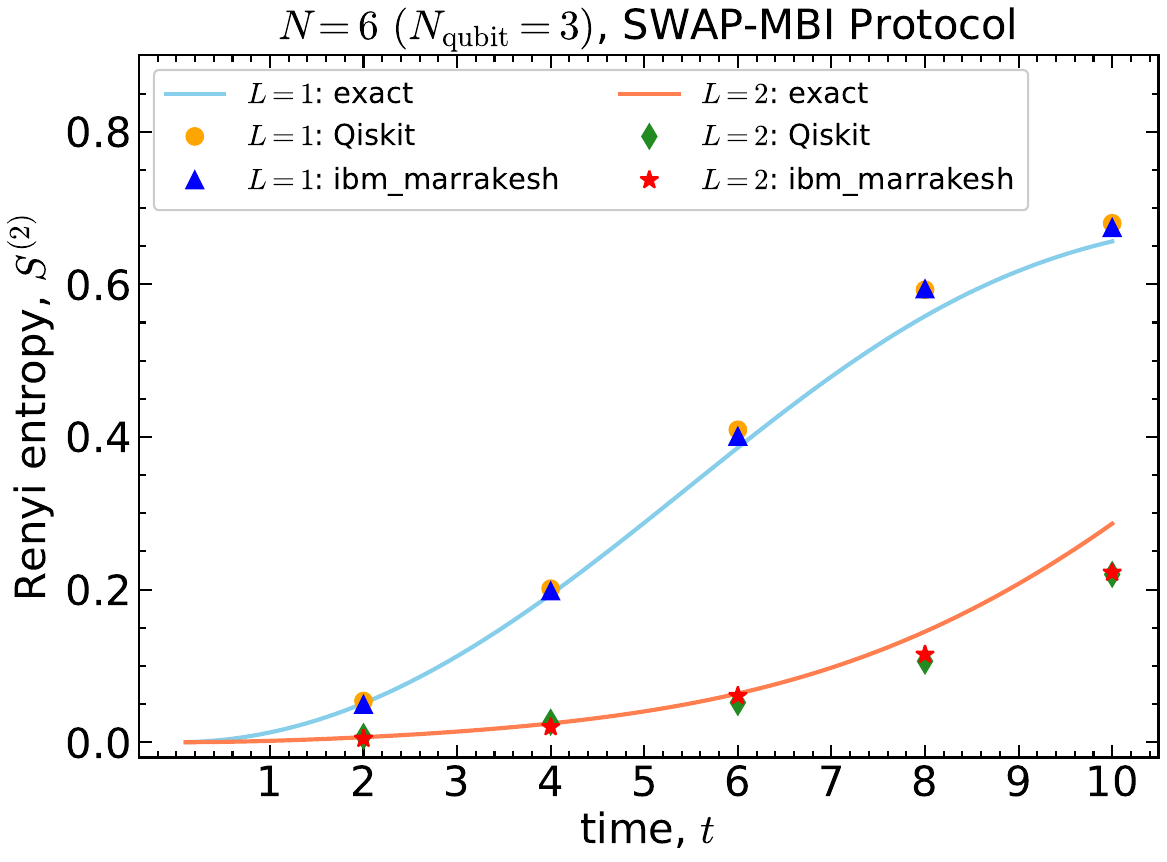}
    \caption{The R\'enyi entropy $S^{(2)}$ with respect to time $t$ (in arbitrary unit) for the left-subsystem of $L=1$ and $L=2$ qubits (two and four Majorana fermions, respectively), determined using swap-based many-body interference protocol (SWAP-MBI). Here, we set the Trotter stepsize as $\delta t = 2$.}
    \label{fig:swap-result}
\end{figure}
During our measurement of the subsystem entanglement entropy over time, we discovered that the SWAP-MBI protocol is more accurate for measuring entropy at early times when it is relatively small. In contrast, the RM protocol, which relies on local Haar-random unitaries, tends to produce incorrect results due to significant statistical fluctuations in its purity estimation during these early times because when the time $t$ is small the quantum state is close to the initial product state, so the R\'enyi entropy associated with its subsystem is relatively small (i.e. the purity is close to 1), and such small values of the R\'enyi entropy are prone to larger statistical errors as pointed out in~\cite{Enk-Beenakker, Elben-2}.

As illustrated in Fig. \ref{fig:swap-result}, there is excellent agreement between the exact and measured R\'enyi entropy $S^{(2)}$ on the \texttt{ibm$\_$marrakesh} device. The slight deviation observed at larger times (specifically, $t = 6-10$) can be attributed to inherent Trotter error. However, at these times, the noiseless Qiskit simulation and the error-mitigated measured values of the SWAP-MBI protocol align very well.
\begin{figure}
    \centering
    \includegraphics[width=1.0\linewidth]{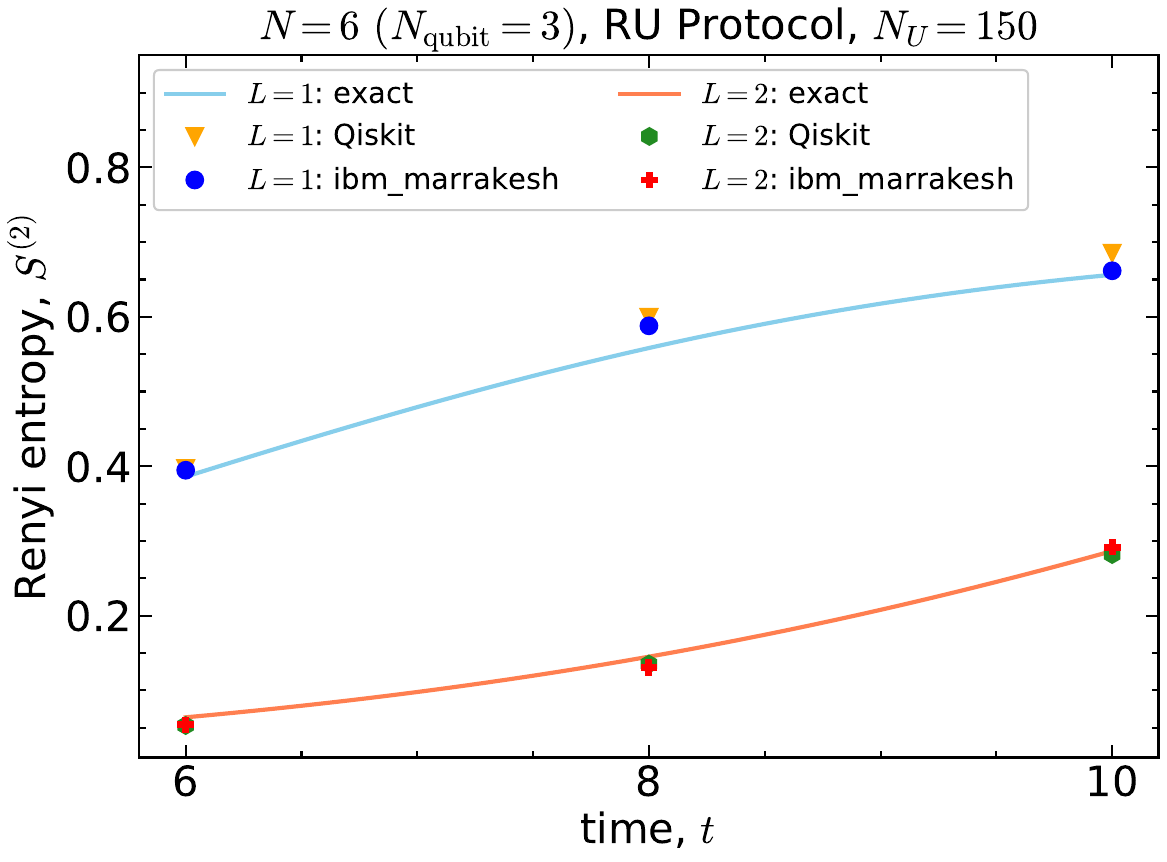}
    \caption{The R\'enyi entropy $S^{(2)}$ with respect to time $t$ determined using the randomized measurement protocol.}
    \label{fig:randomized-result}
\end{figure}

We measure the Rényi entropy using the RM protocol for \( t = 6, 8, \) and \( 10 \), as shown in Fig. \ref{fig:randomized-result}. Again, we see excellent agreement between exact results and results obtained by RM protocol at \texttt{ibm$\_$marrakesh}. A key challenge of the RM protocol is its reliance on an ensemble of local Haar-random unitaries to minimize the variance in entropy computation. In our experiments, we utilize  $150$ samples of the RM protocol circuit. Besides, as we apply the Zero Noise Extrapolation (ZNE) and the Pauli Twirling (PT) methods to mitigate errors, each method replicates the variation three times (ZNE) and ten times (PT), respectively. Therefore, the total number of circuits for the RM protocol at each measurement time is 4,500 ($ 150 \times 3 \times 10$). Since we have three measurement points (6, 8, and 10), the total number of circuits is 135,000.

Therefore, to efficiently utilize the IBMQ computer, we parallelize the execution of quantum circuits through Quantum Multi-Programming (QMP), which allows for the concurrent execution of multiple quantum circuits, regardless of their types or depths of the circuits. The efficiency of the QMP has been demonstrated in various quantum algorithms, such as Grover's search, quantum amplitude estimation, and quantum support vector machine \cite{park2023quantum, rao2024quantum, baker2024parallel}. Despite the efficiency of the QMP, it does introduce some challenges, such as crosstalk between different circuits. We mitigate these side effects following the technique detailed in Ref. \cite{park2023quantum}. In our QMP implementation, we parallelized five circuits within a QMP package with one physical idle qubit between circuits. Therefore, we executed 27,000 circuits by parallelizing them by 5 circuits to execute the 135,000 circuits.

\section{Conclusion and Outlook}\label{sec:conclusion}

Using IBM's superconducting quantum computer, we have obtained the measurement of entanglement entropy growth in the SYK model, which are in excellent agreement with the theoretical results. Despite the challenges posed by the dense interaction structure of the SYK Hamiltonian and the computational demands of entanglement entropy protocols due to the current hardware limitations, we successfully implemented both the SWAP-based many-body interference (SWAP-MBI) and randomized measurement (RM) protocols on IBM's superconducting quantum hardware. To address the limitations of qubit connectivity, we employed an optimized version of the SWAP-MBI protocol and extracted meaningful entanglement dynamics from quantum circuits of large circuit depths.
Importantly, we tackled the substantial resource demands of the RM protocol by utilizing Quantum Multi-Programming (QMP) to parallelize 135,000 circuits. This approach significantly enhanced the efficiency of entropy estimation without compromising accuracy. Finally, our results show excellent agreement with noiseless simulations, validating the SWAP-MBI and RM protocols across different entanglement regimes and timescales. Moreover, the von Neumann entropy in the SYK model can be computed using superconducting quantum computers following~\cite{Xu:2017mvh, Liu:2025tys}.

This work provides a blueprint for measuring entanglement entropy in strongly interacting models on near-term devices and demonstrates how tailored algorithmic and architectural innovations can reveal complex quantum dynamics within the constraints of current quantum hardware. Our approach lays the groundwork for exploring information scrambling, quantum chaos, and holographic dualities in increasingly complex quantum systems.

\section*{Acknowledgments}
T.A.C would like to thank the High Energy Theory group in the Department of Physics and Astronomy at the University of Kansas for their hospitality and support where this work has been done. R.S.S. is supported by Laboratory Directed Research and Development (LDRD No. 23-051) of Brookhaven National Laboratory and RIKEN-BNL
Research Center. 
This work is supported by  the Brookhaven National Laboratory LDRD No. 24-061 (K.Y.).
This research used quantum computing resources of the Oak Ridge Leadership Computing Facility, which is a DOE Office of Science User Facility supported under Contract DE-AC05-00OR22725. 
This research used resources of the National Energy Research Scientific Computing Center, a DOE Office of Science User Facility supported by the Office of Science of the U.S. Department of Energy under Contract No. DE-AC02-05CH11231 using NERSC award DDR-ERCAP0028999 and DDR-ERCAP0033558.



\end{document}